\documentclass[12pt,preprint]{aastex}
\slugcomment{ApJ, in press}

\shorttitle{An Andromedean Triplet}
\shortauthors{McConnachie et al.}

\begin{document}

\title{A trio of new Local Group galaxies with extreme properties}

\author{Alan W. McConnachie$^1$\email{alan@uvic.ca}}
\author{Avon Huxor$^{2}$\email{huxor@roe.ac.uk}}
\author{Nicolas F. Martin$^{3}$\email{martin@mpia-hd.mpg.de}}
\author{Mike J. Irwin$^{4}$\email{mike@ast.cam.ac.uk}}
\author{Scott C. Chapman$^{4}$\email{schapman@ast.cam.ac.uk}}
\author{Gregory Fahlman$^{5}$\email{greg.fahlman@nrc-cnrc.gc.ca}}
\author{Annette M. N. Ferguson$^{2}$\email{ferguson@roe.ac.uk}}
\author{Rodrigo A. Ibata$^{6}$\email{ibata@astro.u-strasbg.fr}}
\author{Geraint F. Lewis$^{7}$\email{gfl@Physics.usyd.edu.au}}
\author{Harvey Richer$^{8}$\email{richer@phas.ubc.ca}}
\author{Nial R. Tanvir$^{9}$\email{nrt3@star.le.ac.uk}}

\affil{$^1$Department of Physics and Astronomy, University of Victoria, Victoria, B.C., V8P 1A1, Canada}
\affil{$^2$Institute for Astronomy, University of Edinburgh, Royal Observatory, Blackford Hill, Edinburgh EH9 3HJ, U.K.}
\affil{$^3$Max-Planck-Institut f{\"u}r Astronomie, K{\"o}nigstuhl 17, D-69117 Heidelberg, Germany}
\affil{$^4$Institute of Astronomy, Madingley  Road,  Cambridge, CB3  0HA,  U.K.}
\affil{$^5$NRC Herzberg Institute of Astrophysics, 5071 West Saanich Road, Victoria, B.C., V9E 2E7, Canada}  
\affil{$^6$Observatoire   de  Strasbourg,   11,   rue  de l'Universite,   F-67000,  Strasbourg,  France}
\affil{$^7$Institute  of Astronomy, School  of Physics, A29, University of  Sydney, NSW 2006, Australia}
\affil{$^8$Department of Physics and Astronomy, University of British Columbia, Vancouver, B.C., V6T 1Z1, Canada} 
\affil{$^9$Department of Physics and Astronomy, University of Leicester, Leicester, LE1 7RH, U.K.\\}

\begin{abstract}
We report on the discovery of three new dwarf galaxies in the Local
Group. These galaxies are found in new CFHT/MegaPrime $g,i$ imaging of
the south-western quadrant of M31, extending our extant survey area to
include the majority of the southern hemisphere of M31's halo out to
150\,kpc. All these galaxies have stellar populations which appear
typical of dwarf spheroidal (dSph) systems. The first of these
galaxies, Andromeda~XVIII, is the most distant Local Group dwarf
discovered in recent years, at $\sim 1.4$\,Mpc from the Milky Way
($\sim 600$\,kpc from M31). The second galaxy, Andromeda~XIX, a
satellite of M31, is the most extended dwarf galaxy known in the Local
Group, with a half-light radius of $r_h \sim 1.7$\,kpc. This is
approximately an order of magnitude larger than the typical half-light
radius of many Milky Way dSphs, and reinforces the difference in scale
sizes seen between the Milky Way and M31 dSphs (such that the M31
dwarfs are generally more extended than their Milky Way
counterparts). The third galaxy, Andromeda~XX, is one of the faintest
galaxies so far discovered in the vicinity of M31, with an absolute
magnitude of order $M_V \sim -6.3$. Andromeda~XVIII, XIX and XX
highlight different aspects of, and raise important questions
regarding, the formation and evolution of galaxies at the extreme
faint-end of the luminosity function. These findings indicate that we
have not yet sampled the full parameter space occupied by dwarf
galaxies, although this is an essential pre-requisite for successfully
and consistently linking these systems to the predicted cosmological
dark matter sub-structure.
\end{abstract}


\section{Introduction}

Edwin Hubble first coined the term ``Local Group'' in his 1936 book
``The Realm of the Nebulae'', to describe those galaxies that were
isolated in the general field but were in the vicinity of the
Galaxy. In recent years, the galaxies of the Local Group have been at
the focus of intense and broad-ranging research, from providing
laboratories for the investigation of dark matter properties (e.g.,
\citealt{gilmore2007} and references therein) to determinations of the
star formation history of the Universe (e.g., \citealt{skillman2005}
and references therein). Understanding individual galaxies in the Local
Group offers important contributions to galaxy structure and evolution
studies; understanding the properties of the population is central to
galaxy formation in a cosmological context.

Hubble originally identified nine members of the Local Group: the
Galaxy and the Large and Small Magellanic Clouds; M31, M32 and NGC205;
M33, NGC6822 and IC1613; along with three possible members NGC6946,
IC10 and IC342. The distances of the latter three were highly
uncertain due to heavy extinction; IC10 has since been confirmed as a
member (\citealt{sakai1999}) although the other two lie outside the
Local Group (NGC6946; \citealt{sharina1997}; IC342:
\citealt{krismer1995}). 

The discovery of new Local Group members continued at a relatively
constant rate up to the start of 2004 (e.g.,
\citealt{ibata1994,whiting1997,whiting1999,
armandroff1998,armandroff1999,karachentsev1999}), at which point the
discovery rate has increased sharply. This has mostly been due to
large area photometric CCD-based surveys of the Milky Way and M31 stellar
haloes: by searching for overdensities of resolved stars in certain
regions of colour-magnitude space, it is possible to identify very
faint dwarf satellites which have previously eluded detection. 

Around the Milky Way, this technique has so far lead to the discovery
of 9 new satellites since 2005 (including possible diffuse star
clusters)
(\citealt{willman2005,willman2006,belokurov2006,belokurov2007,zucker2006b,walsh2007}). All
of these discoveries have been made using the Sloan Digitized Sky
Survey (SDSS). In addition, two new isolated dwarf galaxies have been
identified: Leo T, more than 400kpc from the Milky Way
(\citealt{irwin2007}), was discovered in the SDSS, and a revised
distance estimate for the previously known UGC4879 has moved this
galaxy from $> 10$\,Mpc to being placed on the periphery of the Local
Group (a scant $\sim 1.1$\,Mpc from the Milky Way;
\citealt{kopylov2008}).

Around M31, 9 new dwarf galaxy satellites have been discovered since
2004 (not including results presented herein). Two of these galaxies
(Andromeda IX, X) were found in special SDSS scans of M31
(\citealt{zucker2004a,zucker2007}) and one (Andromeda~XIV) was
discovered serendipitously by \cite{majewski2007} in Kitt Peak 4m imaging of fields in
the south-east halo of M31. The remaining new dwarf galaxies have been
discovered as part of our ongoing photometric survey of this galaxy
and its environs using the INT/WFC (Andromeda~XVII,
\citealt{irwin2008}) and CFHT/MegaPrime (Andromeda~XI, XII and XIII,
\citealt{martin2006}; Andromeda XV and XVI,
\citealt{ibata2007}). Despite its name, Andromeda~XVII is only the
fifteenth dwarf spheroidal satellite of M31 to be discovered;
Andromeda~IV is a background galaxy (\citealt{ferguson2000}) and
Andromeda~VIII was originally identified using planetary nebulae
(\citealt{morrison2003}) which were later shown to belong to M31 and
not to a separate entity (\citealt{merrett2006}). Additionally, only
thirteen of these dwarfs are actually located in the constellation of
Andromeda (Andromeda~VI $\equiv$ the Pegasus dSph; Andromeda~VII
$\equiv$ the Cassiopeia dSph).

The unique, panoramic, perspective of the resolved stellar populations
of galaxies provided by Local Group members make them ideal targets
for observational programs aimed at understanding the detailed
structure of galaxies, their formation processes and their
evolutionary pathways. Dwarf galaxies are of particular interest,
given that they are thought to be the lowest mass, most dark matter
dominated systems which contain baryons (e.g.,
\citealt{mateo1998a}). They are therefore particularly sensitive
probes of external processes, such as tides and ram pressure stripping
(e.g., \citealt{mayer2006,mcconnachie2007c,penarrubia2008b}), and
internal processes such as feedback from star formation (e.g.,
\citealt{dekel1986,dekel2003}). Further, their potential as probes of
dark matter (e.g., \citealt{gilmore2007,strigari2007a}) and their
probable connection to cosmological sub-structures (e.g.,
\citealt{moore1999,bullock2000,kravtsov2004,penarrubia2008a}) give
them an importance to galaxy formation not at all in proportion to
their luminosity.

Here we report on the discovery of three new dwarf galaxies in the
Local Group, all of which have been found as part of our ongoing
CFHT/MegaPrime photometric survey of M31. This new imaging extends our
survey area from the south-eastern quadrant discussed in
\cite{ibata2007} to the west, and currently includes an additional
$49$\,sq.\,degrees of M31's halo out to a maximum projected radius of
150\,kpc. Section~2 summarises the observations and data-reduction
procedures and Section~3 presents a preliminary analysis of the new
dwarfs and quantifies their global properties. In Section~4, we
discuss our results in relation to some of the key questions which
have been prompted with the discoveries of so many new low luminosity
galaxies in the Local Group. Section~5 summarises our results.

\section{Observations}

\cite{martin2006} and \cite{ibata2007} presented first results from our
CFHT/MegaPrime survey of the south-west quadrant of M31, obtained in
semesters S02B -- 06B. Since S06B, we have initiated an extension
to this survey with the aim of obtaining complete coverage of the
southern hemisphere of M31's halo out to a maximum projected radius of
150\,kpc from the center of M31. Figure~1 shows the locations of these
new fields relative to M31 in a tangent-plane projection. Red hatched
fields represent those fields previously presented in
\cite{ibata2007}. Magenta open fields represent the new survey area,
where solid lines denote fields which were observed in S06B -- 07B,
and dotted lines denote fields yet to be observed. Black stars mark
the positions of known M31 satellite galaxies, and open stars mark the
positions of the three new dwarfs presented herein.

Our observing strategy is very similar to that described in
\cite{ibata2007}, to which we refer the reader for further details. In brief,
CFHT/MegaPrime consists of a mosaic of thirty-six $2048 \times
4612$\,pixel CCDs with a total field of view of $0.96 \times
0.94$ sq.\,degrees at a pixel scale of $0.187$\,arcsec\,pixel$^{-1}$. We
observe in the CFHT $g$ and $i$ bands for a total of 1350 seconds
each, split into $3 \times 450$\,seconds dithered sub-exposures, in $<
0.8$\,arcsec seeing. This is sufficient to reach $g \sim 25.5$ and $i
\sim 24.5$ with a signal-to-noise of 10. In some cases, more than
three exposures were taken (at the discretion of CFHT staff to ensure
the requested observing conditions were met), and in these cases the
viable images were included in the stacking procedure, weighted
according to noise/seeing. We have chosen a tiling pattern which
typically has very little overlap between fields, and so we use short,
45\,second exposures in $g$ and $i$ offset by half a degree in the right
ascension and declination directions in order to establish a
consistent photometric level over the survey. This typically has a rms
scatter of 0.02\,mags over our survey area.

The CFHT/MegaPrime data were pre-processed by CFHT staff using the
{\it Elixir} pipeline, which accomplishes the bias, flat, and fringe
corrections and also determines the photometric zero point of the
observations. These images were then processed using a version of the
CASU photometry pipeline (\citealt{irwin2001}) adapted for
CFHT/MegaPrime observations. The pipeline includes re-registration,
stacking, catalogue generation and object morphological
classification, and creates band-merged $g,i$ products for use in the
subsequent analysis.  The CFHT $g$ and $i$ magnitudes are de-reddened
using the \cite{schlegel1998} IRAS maps, such that $g_0 = g - 3.793
E(B - V)$ and $i_0 = i - 2.086 E(B - V)$, where $g_0$ and $i_0$ are
the de-reddened magnitudes.

\section{Analysis}

In this section we present an initial analysis of the three new dwarf
galaxies using the CFHT/MegaPrime discovery data. The measured
parameters of the dwarfs are summarised in Table~1.

\subsection{Discovery and stellar populations}

Two of the new dwarf galaxies (Andromeda XVIII and XIX) stand out as
prominent overdensities of stars in our survey and can be clearly
identified by eye in maps of the distribution of stellar
sources. Andromeda~XX, on the other hand, is considerably fainter and
its CMD is far more sparsely populated. Despite this, it was initially
identified by one of us (A. Huxor) through visual examination of the
individual CCDs during a search for globular clusters. An automated
detection algorithm, based upon a boxcar matched-filter search for
local overdensities with a variable width, was subsequently applied
after these preliminary searches. As well as highlighting these three
dwarfs, some other dwarf galaxy candidates were identified and are
being followed up. A subsequent paper will deal in detail with the
automated detection of dwarf galaxies around M31 to enable a full
completeness study, although such an analysis requires more contiguous
coverage of M31 than we currently possess. Prior to such a study, we
do not make any claims regarding the completeness of the satellite
sample so far discovered.

The top panels of Figure~2 shows the $i_o$ versus $(g - i)_o$
colour-magnitude diagrams (CMDs) for the three new dwarf galaxies
discovered in the south-west quadrant of M31 and whose positions
relative to this galaxy are indicated in Figure~1. The bottom panels
of Figure~2 show reference fields with equivalent areas offset from
the center of each of the galaxies by several half-light radii. Each
of the CMDs has been corrected for foreground extinction. In each of
the three cases, a red giant branch (RGB) is clearly visible, although
in the case of Andromeda~XX it is poorly populated. To the depth of
these observations, it appears that there are very few, if any, bright
main-sequence and blue loop stars which would be indicative of younger
stellar populations, and it is likely therefore that these galaxies do
not host a dominant young population. Stars to the red of the RGB
(with $2 \lesssim (g - i)_o \lesssim 3$) are likely foreground Milky
Way disk stars, although intermediate-age asymptotic giant branch
stars can also occupy this colour locus and have a luminosity similar
to or brighter than the tip of the red giant branch (although this is
probably only relevant for Andromeda~XIX). In the Andromeda~XIX CMD
and reference field, the vertical feature at $(g - i)_o \sim 0.3$ is
the foreground Milky Way halo locus (see \citealt{martin2007} for an
analysis of this feature in our extant M31 survey). Given these
current data, all of the CMDs appear to show stellar populations
typical of dSph galaxies. The faint blue objects centered around $i_o
\sim 25.2$ with a mean colour of $(g - i)_o \sim 0.5$ in the
Andromeda~XIX CMD may be a horizontal branch component. However, as
the reference field shows, contamination from misclassified background
galaxies is considerable in this region of colour - magnitude
space. There is also some evidence of a very weak RGB population in
the Andromeda~XIX reference field, which is likely due to the
background M31 halo and stellar overdensities in the vicinity of
this dwarf galaxy (see Section 4.3.2).

Figure~3 shows various properties for each of the three new dwarf
galaxies. The left-most panels show $I_o$ versus $(V - I)_o$ CMDs for
each galaxy.  We have transformed CFHT $gi$ to Landolt $VI$ using a
two-stage transformation; we first change CFHT $gi$ into INT $V^\prime
i$ using the relations derived in \cite{ibata2007}, and we then
transform INT $V^\prime i$ into Landolt $VI$ using the transformations
given in \cite{mcconnachie2004a}\footnote{see
http://www.ast.cam.ac.uk/$\sim$wfcsur/colours.php for details}. In
each CMD, only those stars which lie within two half-light radii from
the center of each galaxy (shown by the red dashed ellipse in the
second panel) have been plotted. The dashed lines define a colour cut
designed to preferentially select stars which are members of the dwarf
galaxies. The solid line shows a 13\,Gyr isochrone with the
representative metallicity of the dwarf from \cite{vandenberg2006},
shifted to the distance modulus of the dwarf (the distance and
metallicity of each dwarf is calculated in Section~3.2).

The second panel in each row of Figure~3 shows the spatial
distribution of candidate RGB stars in the vicinity of each galaxy,
defined by the colour cuts discussed previously. Black dashed lines
show the edges of the CFHT/MegaPrime CCDs. Both Andromeda~XVIII and XX
appear as obvious concentrations of stars, despite Andromeda~XX being
poorly populated. Andromeda~XVIII lies at the corner of one of the
CCDs, and much of this galaxy hides behind the large gap between the
second and first rows of CCDs in the CFHT/MegaPrime field (see Section
3.3). Andromeda~XIX is a much more extended and diffuse system than
the other two, and contours have been overlaid to more clearly show
its structure. The first contour is set $3 - \sigma$ above the
background, and subsequent contour levels increase by $1.5\,\sigma$
over the previous level. This galaxy is located on the boundary of our
survey, overlapping slightly with the extant survey region from
\cite{ibata2007}. We include some adjacent fields from this earlier
part of the survey to obtain complete coverage of Andromeda~XIX.

\subsection{Distances and metallicities}

The upper-right panels in each row of Figure~3 show, for each galaxy,
the de-reddened $I$-band luminosity functions of stars in the CMD
which satisfy the colour and spatial cuts defined previously. These
have been corrected for foreground/background contamination by
subtracting a nearby ``reference'' field, scaled by area. The scaled
reference field is shown by the dotted line, to illustrate the
contribution from the foreground/background as a function of
magnitude. The $I-$band magnitude of the tip of the red giant branch
(TRGB; corresponding to the point in the evolution of a RGB star
immediately prior to it undergoing the core helium flash) is a
well-calibrated standard candle which is used extensively for nearby
galaxies (e.g.,
\citealt{lee1993a,salaris1997,mcconnachie2004a,mcconnachie2005a} and
references therein). In a well populated luminosity function, it is
normally taken to be equal to the luminosity of the brightest RGB
star. However, when dealing with faint dwarfs - particularly systems
like Andromeda~XX with a very sparse RGB - this assumption is likely
to be flawed due to sampling errors. However, for this initial
analysis of these galaxies we assume that the TRGB position measured
in this way is a good estimate of its actual position. We note that
the resulting distance modulus of Andromeda~XX in particular is
uncertain and will be refined once deeper data reaching below the
horizontal branch is available.

Our best estimates for the (extinction-corrected) $I-$band magnitude
of the TRGB are highlighted on each of the luminosity functions in
Figure~3 and are listed in Table~1. For Andromeda~XX, we have adopted
very conservative error bars; the lower limit is an estimate of the
possible offset of the brightest RGB star from the true TRGB from our
experience with the comparably faint Andromeda~XII (\citealt{chapman2007}); the upper limit
assumes that the few brightest stars we have identified are actually
foreground contamination, and that the true TRGB is represented by the
group of stars at $I_o \sim 21.2$.  Adopting $M_I = -4.04 \pm 0.12$
(\citealt{bellazzini2001}) yields a preliminary distance to each of
the new dwarf galaxies; the derived distance moduli and distances are
given in Table~1. Most notable is the distance to Andromeda~XVIII,
which has a well-defined TRGB, and which places it approximately
1.4\,Mpc from the Milky Way ($\sim 600$\,kpc distant from M31), at the
periphery of the Local Group.

As an independent check of our distance estimates (particularly that
for Andromeda~XX), we construct $g_o$ luminosity functions for each
galaxy using stars within two half-light radii of the centers. These
are shown in Figure~4 as solid lines. Also shown as dotted lines are
luminosity functions for nearby reference fields, scaled by
area. These luminosity functions go deeper than the previous CMDs
since stars are only required to be detected in the $g$-band. Our data
start to become seriously incomplete below $g_o \sim 25.5$, and
photometric errors at this magnitude are of order $\Delta\,g \simeq
0.15$.  For reference, the horizontal branch in M31 has a magnitude of
$g_o \sim 25.2$ (\citealt{ibata2007}). For Andromeda~XIX and XX, peaks
of stars are visible at $g \sim 25.3$ and $g \sim 25.6$, respectively,
which are not present in the reference fields, and which are marked in
Figure~4 by dashed lines. We attribute these peaks to the detection of
horizontal branch stars in each of these galaxies. While the peak for
Andromeda~XIX is less apparent than that for Andromeda~XX, its
position coincides with the expected luminosity of the horizontal
branch from inspection of its CMD in Figure~2, reinforcing our
interpretation of this feature. In contrast, no such feature is
visible for Andromeda~XVIII, which is expected given that we measure
it to be much more distant than the other two and so our observations
will not be deep enough to observe the horizontal branch
population. Similarly, our measurements of the positions of the
horizontal branches in Andromeda~XIX and XX are consistent with the
positions we measure for the TRGB in these galaxies.  These detections
(and non-detection) of the horizontal branches are therefore
consistent with the distances derived from the TRGB, and suggest that
the uncertainty in the distance to Andromeda~XX may be less than we
currently adopt in Table~1.

The lower-right panels of Figure~3 show the observed photometric
metallicity distribution (MDF) function, constructed using the same
technique as detailed in \cite{mcconnachie2005a}, using a bi-linear
interpolation of stars in the top two magnitudes of the RGB with
13\,Gyr isochrones, [$\alpha$/Fe] $= 0$, from \cite{vandenberg2006}
with $BVRI$ colour-$T_{eff}$ relations as described by
\cite{vandenberg2003}. Each MDF has been corrected for
foreground/background contamination by subtraction of a MDF for a
reference field, scaled by area. The MDF for the scaled reference
field is shown as a dotted line in each panel. The mean metallicity
and metallicity spread, as quantified by the inter-quartile range
(IQR), are highlighted in Figure~3, and an isochrone corresponding to
the mean metallicity of the dwarf is overlaid on the CMD in the first
panels, shifted to the distance modulus of the dwarf galaxy.

The metallicity spread in each of the three galaxies is similar,
although the IQR for Andromeda~XIX appears slightly smaller than for
the other two. Certainly, the colour spread of the RGB seen from the
CMDs is much smaller for Andromeda~XIX than for Andromeda~XVIII and
XX. That this does not correspond to a much smaller spread in
metallicity probably reflects the metal poor nature of Andromeda~XIX,
since RGB colour is a poor indicator of metallicity variation at very
low metallicities. It is also tempting to suggest that the narrow
spread in RGB colour indicates that Andromeda~XIX is a simple stellar
population; however, lessons learned from the Carina dSph, which has a
large age and metallicity spread but conspires to have a narrow RGB
(\citealt{smeckerhane1994}), suggests a note of caution against this
interpretation.

The metallicity information is summarised in Table~1. The formal
uncertainties in the metallicity and metallicity spread estimates are
of order 0.1\,dex. In addition to uncertainties in the stellar models,
our metallicity estimates assume that (i) the dwarfs are all dominated
by a $13$\,Gyr stellar population, and (ii) the distance modulus for
each galaxy is well estimated. The former assumption is likely
reasonable, and should not lead to an error $\gtrsim 0.2$\,dex unless
the dwarfs are dominated by intermediate-age and young stellar
populations (for which there is no current evidence). The latter
assumption looks to be reasonable for Andromeda~XVIII and XIX, where
the RGB is reasonably well populated, but for Andromeda~XX the
uncertainty introduced through the distance estimate could be more
significant. We note that the metallicities of Andromeda~XVIII and XIX
look to be significantly lower than the median metallicity of the
kinematically-selected halo of M31, which has [Fe/H]$ \simeq -1.4$
(\citealt{chapman2006,kalirai2006b}).

\subsection{Structures and magnitudes}

We quantify the structures of Andromeda~XVIII, XIX and XX through the
spatial distributions of their resolved stars. However, the analysis
is made more complex since Andromeda~XIX is very diffuse, Andromeda~XX
has very few bright stars on which to base our analysis, and each of
the dwarf galaxies lies close to or at the edges of CCDs. In the
extreme case of Andromeda~XVIII, we are clearly missing a significant
part of the galaxy which lies behind the large gap between the second
and first rows of the CFHT/MegaPrime mosaic. To illustrate this, the
top panel of Figure~5 shows the $i-$band image of Andromeda~XVIII with
linear scaling; while Andromeda~XVIII is clearly visible to the naked
eye, much of the galaxy falls off the edge of the detector. To
determine how large this effect is, the lower panel of Figure~5 shows
a $10 \times 10$\,arcmins image centered on the coordinates of
Andromeda~XVIII from the POSSII/UKSTU (Blue) survey which we retrieved
through the Digitized Sky Survey, and which covers the entirety of
this galaxy.

Given the several complications discussed above, we choose to derive
the structural parameters for the dwarfs based upon the maximum
likelihood technique developed by \cite{martin2008} instead of the
usual technique which bins the data spatially and uses smoothing
kernels (e.g., \citealt{irwin1995,mcconnachie2006b}). The procedure
has been modified from \cite{martin2008}, to which we refer the reader
for details, to account for incomplete coverage of the dwarfs due to
CCD edges. In brief, this technique calculates simultaneously the most
plausible values for the centroid, ellipticity, position angle and
half-light radius of the dwarf under the assumption that the surface
brightness radial profile is well described by an exponential curve, without
any need for smoothing or binning of the data. However, for
Andromeda~XVIII this approach is still insufficient since our data
only samples one segment of the galaxy, as shown by comparing the
POSSII/UKSTU image with the CFHT/MegaPrime image in Figure~5. Thus, for this
galaxy, we estimate its center from the POSSII/UKSTU data and
approximate it as circular. The half-light radius is then calculated
via the same technique as for Andromeda~XIX and XX using the CFHT/MegaPrime
data.

The centroid, half-light radius ($r_h$), position angle (measured east
from north) and ellipticity ($\epsilon = 1 - b/a$) for each dwarf
galaxy, derived using the maximum likelihood technique (with the above
caveat for Andromeda~XVIII), are listed in Table~1. In addition,
Figure~6 shows the (background-corrected) stellar density profile
(equivalent to the surface brightness profile), derived using the same
technique as in \cite{mcconnachie2006b}, for each of the three dwarf
galaxies. We use elliptical annuli with the position angle,
ellipticity and centroid listed in Table~1. Overlaid on these profiles are
exponential profiles with the appropriate half light radii (the
exponential scale radius, $r_e \simeq 0.6 r_h$). These profiles are
the most probable exponential models for the stellar density
distribution of the dwarf galaxy derived using the maximum likelihood
method, and are not fits to the averaged data-points.

We estimate the magnitude of Andromeda~XIX and XX in a similar way as
\cite{martin2006} and \cite{ibata2007}. First, we sum the total
$V-$band flux from candidate member stars which are within the
half-light radius of each dwarf galaxy and which are within $2 - 3$
magnitudes of the TRGB.  However, this flux does not take into account
the contribution to the total light from fainter stars, most of which
we do not detect. To determine the appropriate correction to apply, we
compare the half-light flux of Andromeda~III measured in this way
(using similar CFHT/MegaPrime observations) to its apparent magnitude
of $m_v = 14.4 \pm 0.3$, directly measured by
\cite{mcconnachie2006b}. We then apply the appropriate correction to
the fluxes for each dwarf galaxy. Clearly, the uncertainties
associated with this method are considerable, and we make the implicit
assumption that the luminosity functions of Andromeda~III, XIX and XX
are similar. Under this assumption, we estimate an accuracy of $\sim
0.6$\,mags in the final magnitude of Andromeda~XIX, although we
estimate a larger uncertainty of $\sim 0.8$\,mags for Andromeda~XX due
to the small number of bright stars available. The central surface
brightness of Andromeda~XIX and XX are estimated by normalizing the
exponential profiles shown in Figure~6 so that the surface integral
over the dwarf out to the half-light radius is equal to half the total
flux received from the dwarf. These numbers are also given in Table~1.

It is not possible to derive the magnitude of Andromeda~XVIII
in the same way as above given that we only sample a segment of this
galaxy with our data. Comparison of the POSSII/UKSTU images of
Andromeda~XVIII with those of Andromeda~V, VI and VII show that it is
considerably lower surface brightness than either Andromeda~VI or VII,
but is similar to - and perhaps brighter than - that of Andromeda~V,
which has $S_o = 25.6 \pm 0.3$ (\citealt{mcconnachie2006b}). We
therefore adopt this as a faint-end limit to the central surface
brightness of Andromeda~XVIII. A faint-end limit to its magnitude can
then be calculated by normalizing its radial surface brightness
profile to this central value, integrating over its area out to the
half light radius, and multiplying the answer by two. The magnitude
derived in this way is given in Table~1. We note that updated
magnitudes and surface brightnesses will be derived for each of the
three new galaxies using the unresolved light component from
dedicated, follow-up, photometric studies.

\section{Discussion}

Andromeda~XVIII, XIX and XX have a range of relatively unusual
properties. In particular, Andromeda~XVIII is one of the most distant
Local Group galaxies discovered for several years, and is one of the
most isolated systems in the Local Group. Andromeda~XIX is extremely
extended, with a very large half-light radius and extremely faint
central surface brightness. Andromeda~XX, on the other hand, is one of
the lowest luminosity dwarf galaxies so far discovered around M31,
with a magnitude of $M_V \simeq -6.3^{+1.0}_{-0.7}$, comparable to the
luminosity of Andromeda~XII ($M_V = -6.4 \pm 1.0$;
\citealt{martin2006}). In this section, we discuss the properties of
these galaxies in the larger context of the main science questions
raised by the recent discoveries of so many new dwarf galaxies.

\subsection{Completeness}

Prior to 2004, there were 15 dSph galaxies known in the Local Group
(nine Milky Way satellites, six M31 satellites and two isolated
systems, Cetus and Tucana). Since this time, 22 new dwarf galaxies
(including possible diffuse star clusters around the Milky Way) have
been discovered in the Local Group, the overwhelming majority of which
are dSph satellites of the Milky Way and M31. For the Milky Way, the
SDSS has been responsible for all the discoveries to date, and most of
the galaxies discovered have been extremely faint; no new Milky Way
satellites with $M_V \lesssim -8$ have been found. Thus, apart from
satellites hidden by the Milky Way disk, our satellite system is
probably complete to this approximate magnitude limit, as originally
argued by \cite{irwin1994}.

Around M31, it is more difficult to identify extremely faint dwarf
galaxies since we cannot probe as far down the stellar luminosity
function. Andromeda~XII and Andromeda~XX are the two faintest M31
satellites found so far, both with $M_V \sim -6.3$. For comparison,
the faintest Milky Way satellite found to date is probably Willman I,
with $M_V \sim -2.7$ (\citealt{willman2006,martin2008}).

Andromeda~XVIII is considerably brighter than Andromeda~XX, and has a
central surface brightness similar to or brighter than Andromeda~V
($S_o = 25.6 \pm 0.3$\,mags\,arcsec$^{-2}$). Andromeda~XVIII is
clearly visible in the POSSII/UKSTU (Blue) survey image which we
retrieved through the Digitized Sky Survey and which is reproduced in
the lower panel of Figure~5. However, its identification is made more
complicated by numerous nearby bright stars and nebulosity in its
vicinity, which may act to explain why it was not discovered using
these data. We have also confirmed that it is visible in the original
POSSI (Blue) survey. Its belated discovery indicates that previous
surveys for relatively {\it bright} dwarf galaxies around M31 were
incomplete and that some dwarfs were missed. Variable and unknown
completeness is problematic for studies of satellite distributions and
highlights the vital need for more systematic studies such as those
now being conducted.

It is fortuitous that Andromeda~XVIII lies within our survey area
given its considerable distance from M31. Indeed, even as current and
future surveys help improve the completeness of the M31 and Milky Way
satellite systems, many isolated Local Group galaxies can be expected
to continue to elude detection: unlike the Milky Way satellites, they
are not nearby, and unlike the M31 satellites, they are not
necessarily clustered in an area amenable to systematic
searches. PanStarrs $3\pi$ will survey a large fraction of the sky a
magnitude deeper than SDSS, and should discover isolated Local Group
galaxies, particularly those within 500\,kpc or so from the Milky
Way. However, very faint galaxies much further away than this ($\sim
1$\,Mpc) may prove more difficult to spot. Exactly how many very faint
dwarf galaxies are to be found at the periphery of the Local Group is
likely to remain uncertain for some time yet.

\subsection{Spatial distribution}

Several recent studies of the spatial distributions of satellites
around the Milky Way and M31
(\citealt{willman2004,kroupa2005,mcconnachie2006a,koch2006,metz2007,irwin2008})
have generally concluded that the distributions appear anisotropic:
\cite{mcconnachie2006a} highlight the fact that (at the time) 14 out
of the 16 candidate satellites of M31 are probably on the near side of
M31, while others (\citealt{kroupa2005,koch2006,metz2007,irwin2008})
conclude that many of the Milky Way and M31 satellites are aligned in
very flattened, disk-like, distributions (an observation originally
made by \citealt{lyndenbell1976,lyndenbell1982}).

Andromeda~XVIII, XIX and XX do not lie near any of the principle
satellite planes previously proposed to exist around M31. As discussed
in the previous sub-section, the census of Local Group galaxies is
clearly not complete, and it is too early to draw definitive
conclusions regarding the distributions of satellites. This is
particularly true around M31, where relatively bright satellites are
still being discovered. For the Milky Way, the SDSS covers roughly
one-fifth of the Milky Way halo in the direction of the north galactic
cap; depending upon how many satellites are found in future surveys at
lower latitudes, the statistical significance of the proposed streams
of satellites may change substantially.

In terms of spatial distributions, Andromeda~XVIII is unusual insofar
as it is very distant - roughly 1.4\,Mpc from the Milky Way, and
roughly 600\,kpc from M31. Thus it is probably not a satellite of M31,
although kinematics may help reveal whether it is approaching M31 and
the Local Group for the first time (like Andromeda~XII,
\citealt{chapman2007}) or if it has been thrown out from M31 following
an interaction (like Andromeda~XIV, \citealt{majewski2007,sales2007}).

\subsection{Environment and structures}

\subsubsection{Andromeda~XVIII, position and morphology} 

Andromeda~XVIII appears to possess stellar populations typical of dSph
galaxies. If it is subsequently confirmed to be gas poor, then it will
be the third dSph galaxy found in isolation in the Local Group (in
addition to Cetus and Tucana). The fact that isolated galaxies are
preferentially more gas-rich compared to satellites
(\citealt{einasto1974}) has lead to the proposition that satellite
galaxies are stripped of their gas via ram-pressure stripping and
tidal harassment in the halo of the host galaxy (e.g.,
\citealt{mayer2006}). However, for isolated systems such as
Andromeda~XVIII, Cetus and Tucana, prolonged interactions with massive
galaxies are unlikely to have occurred. Likewise, the gas-deficient
satellite Andromeda~XII is not believed to have undergone any past
interactions with a large galaxy since it appears to be on its first
infall into the potential of M31 (\citealt{chapman2007}). Further, the
most compelling case of a dwarf galaxy thought to be undergoing
ram-pressure stripping is Pegasus (DDO216;
\citealt{mcconnachie2007c}), an {\it isolated} galaxy more than
400\,kpc from M31. Clearly, understanding if these observations are
consistent with the present models for dwarf galaxy evolution requires
a more complete inventory of nearby galaxies and their properties than
we currently possess.

\subsubsection{Andromeda~XIX, tides and substructure}
 
The half-light radius of Andromeda~XIX is $6.2$\,arcmins. At the
distance we derive for it, this corresponds to $r_h \simeq 1.7$\,kpc,
which is the largest value yet recorded for any dSph in the Local
Group. The average half-light radius for Milky Way dSphs is an order
of magnitude less, at $r_h \sim 150$\,pc, and none have half-light
radii larger than $r_h \simeq 550$\,pc (with the exception of the
tidally disrupting Sagittarius dSph; \citealt{majewski2003}) . M31
dSphs, on the other hand, have typical half-light radii of $r_h \sim
300$\,pc, with the previous extremes being Andromeda~II, with $r_h
\simeq 1.1$\,kpc, and Andromeda~VII, with $r_h \simeq 750$\,pc
(\citealt{mcconnachie2006b}).  The extremely diffuse and extended
nature of Andromeda~XIX is reminiscent of the ``outer component'' of
Andromeda~II, as traced by horizontal branch stars by
\cite{mcconnachie2007a}.

It is tempting to attribute the diffuse structure of Andromeda~XIX to
tidal interactions. In this respect, it is relevant to note that
Andromeda~XIX lies very close to the major axis substructure
identified by \cite{ibata2007}. No independent distance estimate to
this substructure currently exists; \cite{ibata2007} assumed it to be
at the distance of M31 but if it is at the same distance as
Andromeda~XIX then the photometric metallicity estimates of these
features will be very similar. Figure~7 shows the surroundings of
Andromeda~XIX as a stellar density map; the first two contour levels
are $2$ and $3 - \sigma$ above the background, and the levels then
increase by $1.5\,\sigma$ over the previous level. As well as showing
Andromeda~XIX as a prominent overdensity, there is some evidence of
stellar material in its outskirts (also visible in the contours of
Figure~3). Whether or not Andromeda~XIX is the source of the major
axis substructure identified in \cite{ibata2007}, or is being tidally
perturbed, will require detailed kinematics in this region. We note
that \cite{penarrubia2008b} show that the effect of tides on dwarf
galaxies in cosmological haloes is to decrease the central surface
brightness and {\it decrease} the half light radius of the bound
component. This would argue against tidal effects explaining the
structure of Andromeda~XIX.

The large scale-size of Andromeda~XIX reinforces the difference in
scale-size between the Milky Way and M31 satellites first highlighted in
\cite{mcconnachie2006b}, such that the M31 dSphs are more extended
than their Milky Way counterparts.
\cite{penarrubia2008a,penarrubia2008b} have investigated the cause of
this disparity in an attempt to relate it to either differences in the
underlying dark matter properties of the dwarfs or differences in
their evolution around their hosts. They conclude that tidal effects
are insufficient to explain the magnitude of the effect. However, if
the different scale sizes reflect intrinsic differences between the
Milky Way and M31 sub-haloes then this should reveal itself in the
kinematics of the two populations (with the M31 dwarfs being
dynamically hotter than their Milky Way counterparts). Whatever the
cause, the comparison of Andromeda~XIX and the other M31 satellites to
the Milky Way population highlights the importance of sampling dwarfs
in a range of environments so as to obtain a fuller appreciation of
the range of properties that these systems possess. In turn, this
helps us understand the physical drivers behind the differences and
similarities we observe. We note that studies of the star clusters of
M31 (\citealt{huxor2005,huxor2008}) have already extended the known
parameter space for these objects, with the M31 population containing
extended star clusters not found in the Milky Way population.

\subsection{Satellites that are missing and ``the missing satellites''}

Andromeda~XX is an exceptionally faint galaxy with a very poorly
populated RGB. This makes an accurate derivation of its properties
particularly difficult. However, the star formation history of
Andromeda~XX and the other ultra-faint satellites is particularly
relevant to the ``missing satellites'' question (do all the thousands
of dark matter sub-haloes predicted to exist in the haloes of galaxies
like the Milky Way and M31 contain stars and, if they do, where are
they?).  Until recently, only a dozen or so dwarf satellites were
observed, and it was noted that the cumulative mass distribution of
these satellites was dramatically different to that of predicted dark
matter sub-haloes, even at relatively large masses
(\citealt{moore1999,klypin1999}). To solve this discrepancy without
altering the underlying cosmology, it was suggested that either there
were a large number of luminous satellites awaiting discovery or that
not all sub-haloes have a luminous component.

Despite many new galaxies in the Local Group being discovered, and
many more undoubtedly awaiting discovery, we consider it very unlikely
that these discoveries will resolve the discrepancy between theory and
observation. The original comparison between the observed and
predicted satellite mass functions shows that the discrepancy sets in
for dwarfs as luminous as the Small Magellanic Cloud ($M_V \simeq
-16$) and Fornax ($M_V \simeq -13 $). Finding thousands of very faint
(and presumably less massive?) satellites would not solve the
disagreement at the more massive end and there is no evidence to
suggest that a dozen galaxies the luminosity of Fornax have been
missed (e.g., \citealt{irwin1994}). Further, as higher resolution dark
matter simulations make clear (e.g., \citealt{diemand2007}), the
sub-halo mass function appears to continue to increase at the low mass
end. It seems reasonable, therefore, that at some point these haloes
will not be massive enough to be able to accrete and/or retain baryons
and form stars, and this implies that there is a minimum mass halo
which can host a luminous component (\citealt{kravtsov2004}).

A re-analysis of the observed dynamics of the dwarf galaxies by
\cite{penarrubia2008a,penarrubia2008b} within the $\Lambda-$CDM
çframework has shown that few-if-any of these galaxies (including
recent discoveries) occupy a halo with a circular velocity less than
$\sim 10 - 20$\,km\,s$^{-1}$. Further, these estimates bring the
cumulative distribution of luminous satellites and dark matter
sub-haloes into good agreement at the high-mass end. Using a different
technique, \cite{strigari2007b} find a similar result. Given that
these authors find good agreement between observations and theory down
to a certain mass limit, their results support the idea of a mass
threshold in dark matter haloes below which star formation becomes
highly inefficient. Therefore, by continuing to identify new,
ultra-faint dwarfs, we probe the astrophysics of galaxy formation at
low mass limits where the sensitivity to complex feedback mechanisms -
such as star formation (\citealt{kravtsov2004}) and reionization
(\citealt{bullock2001}) - is greatest.

\section{Summary}

We have presented three new Local Group dwarf galaxies discovered as
part of our ongoing CFHT/MegaPrime survey of M31 and its
environs. These galaxies - christened Andromeda~XVIII, XIX and XX
after the constellation in which they are found - have stellar
populations which appear typical of dSph galaxies. Individually, each
of these galaxies has relatively unusual properties compared to the
previously known dwarfs in the vicinity of M31:

\begin{itemize}
\item Andromeda~XVIII is extremely distant, at $1355 \pm 88$\,kpc from
the Milky Way, placing it nearly $600$\,kpc from M31. Thus it is one
of the most isolated galaxies in the Local Group. It is clearly
observed through its integrated light (it appears to have a central
surface brightness similar to or brighter than that of Andromeda~V)
and suggests that there could be several other relatively bright dwarf
galaxies within the Local Group which have so far eluded detection;

\item Andromeda~XIX is extremely extended, with a half-light radius of
$r_h =  1683 \pm 113$\,kpc. This is an order of magnitude more extended
than typical Milky Way dSphs. While its integrated luminosity is  $M_V = -9.3 \pm 0.6$, its central surface brightness is exceptionally
low, at $S_o = 29.3 \pm 0.7$. Andromeda~XIX reinforces the difference in
scale-size between the Milky Way and M31 satellites first discussed in
\cite{mcconnachie2006b}. This galaxy may be being tidally disrupted,
and could be related to major axis substructure first identified in
\cite{ibata2007} and which lies near to Andromeda~XIX in
projection. However, we note that calculations by
\cite{penarrubia2008b} show that the net effect of tides on a dwarf
galaxy is to decrease the central surface brightness and {\it
decrease} the half-light radius of the bound component;

\item Andromeda~XX is extremely faint, with an absolute magnitude of
order $M_V = -6.3^{+1.0}_{-0.7}$. It is one of the faintest galaxy so far
discovered in the vicinity of M31 (comparable in luminosity to
Andromeda XII) and as such many of its key parameters are extremely
uncertain at this stage. A full inventory of these systems is required
to properly define the faint-end of the galaxy luminosity function,
and to determine where, if anywhere, we encounter a lower limit to the
galaxy mass/luminosity function.
\end{itemize}

\acknowledgements Based on observations obtained with
MegaPrime/MegaCam, a joint project of CFHT and CEA/DAPNIA, at the
Canada-France-Hawaii Telescope (CFHT) which is operated by the
National Research Council (NRC) of Canada, the Institute National des
Sciences de l'Univers of the Centre National de la Recherche
Scientifique of France, and the University of Hawaii. We are indebted
to the CFHT staff for their help and careful observations, and we
thank the anonymous referee for useful comments which improved the
clarity of this paper. AWM thanks Evan Skillman, Jorge Pe{\~n}arrubia
and Andrew Cole for useful discussions. AWM is supported by a Research
Fellowship from the Royal Commission for the Exhibition of 1851, and
thanks Sara Ellision and Julio Navarro for additional financial
assistance. AH and AMNF acknowledge support from a Marie Curie
Excellence Grant from the European Commission under contract
MCEXT-CT-2005-025869.

\clearpage

\begin{table*}
\begin{center}
\begin{tabular}{lccc}
\hline
                 & Andromeda XVIII                                  & Andromeda XIX                                    & Andromeda XX\\
\hline
$\alpha$ (J2000) & 00\,h\,02\,m\,14.5\,s~~($\pm 10''$)              & 00\,h\,19\,m\,32.1\,s~~($\pm 10''$)              & 00\,h\,07\,m\,30.7\,s~~($\pm 15''$) \\
$\delta$ (J2000) & +45$^\circ$\,05$^\prime$\,20$''$ ($\pm 10''$)    & +35$^\circ$\,02$^\prime$\,37.1$''$~~($\pm 10''$) & +35$^\circ$\,07$^\prime$\,56.4$''$~~($\pm 15''$) \\
$(l, b)$         & $(113.9^\circ, -16.9^\circ)$                     & $(115.6^\circ, -27.4^\circ)$                     & $(112.9^\circ, -26.9^\circ)$\\
$E(B - V)$       & 0.104                                            & 0.062                                            & 0.058\\
$I_{o,trgb}$     & $21.62 \pm 0.05$                                 & $20.81 \pm 0.05$                                 & $20.48^{+0.73}_{-0.20}$\\
$(m - M)_o$      & $25.66 \pm 0.13$                                 & $24.85 \pm 0.13$                                 & $24.52^{+0.74}_{-0.24}$\\   
Distance         & $1355 \pm 88$\,kpc                               & $933 \pm 61$\,kpc                                & $802^{+297}_{-96}$\,kpc\\
$r_{M31}$        & $\sim 589$\,kpc                                  & $\sim 187$\,kpc                                  & $\sim 129$\,kpc\\
${\rm [Fe/H]}$   & $-1.8 \pm 0.1$                                   & $-1.9 \pm 0.1$                                   & $-1.5 \pm 0.1$\\
$IQR$            & $0.5$                                            & $0.4$                                            & $0.5$\\
$r_h$ (arcmins)  & $0.92^{+0.05}_{-0.06}$                           & $6.2 \pm 0.1$                                    & $0.53^{+0.14}_{-0.04}$\\
$r_h$ (pc)       & $363^{+31}_{-33}$                                & $1683 \pm 113$                                   & $124^{+56}_{-18}$\\
PA (N to E)      & $0$                                              & $(37^{+4}_{-8})^\circ$                           & $(80 \pm 20)^\circ$\\
$\epsilon = 1 - b/a$ & $0$                                          & $0.17 \pm 0.02$                                  & $0.3 \pm 0.15$\\
$m_v$            & $\le 16.0$                                         & $15.6 \pm 0.6$                                   & $18.2 \pm 0.8$ \\
$M_V$            & $\le -9.7$                                       & $-9.3 \pm 0.6$                                   & $-6.3^{+1.0}_{-0.8}$\\
$S_o$            & $\le 25.6$                                       & $29.3 \pm 0.7$                                   & $26.2 \pm 0.8$\\
\hline
\end{tabular}
\end{center}
\caption{Properties of Andromeda XVIII, XIX and XX}
\end{table*}

\clearpage

\begin{figure}
  \begin{center}
    \includegraphics[angle=270, width=14.cm]{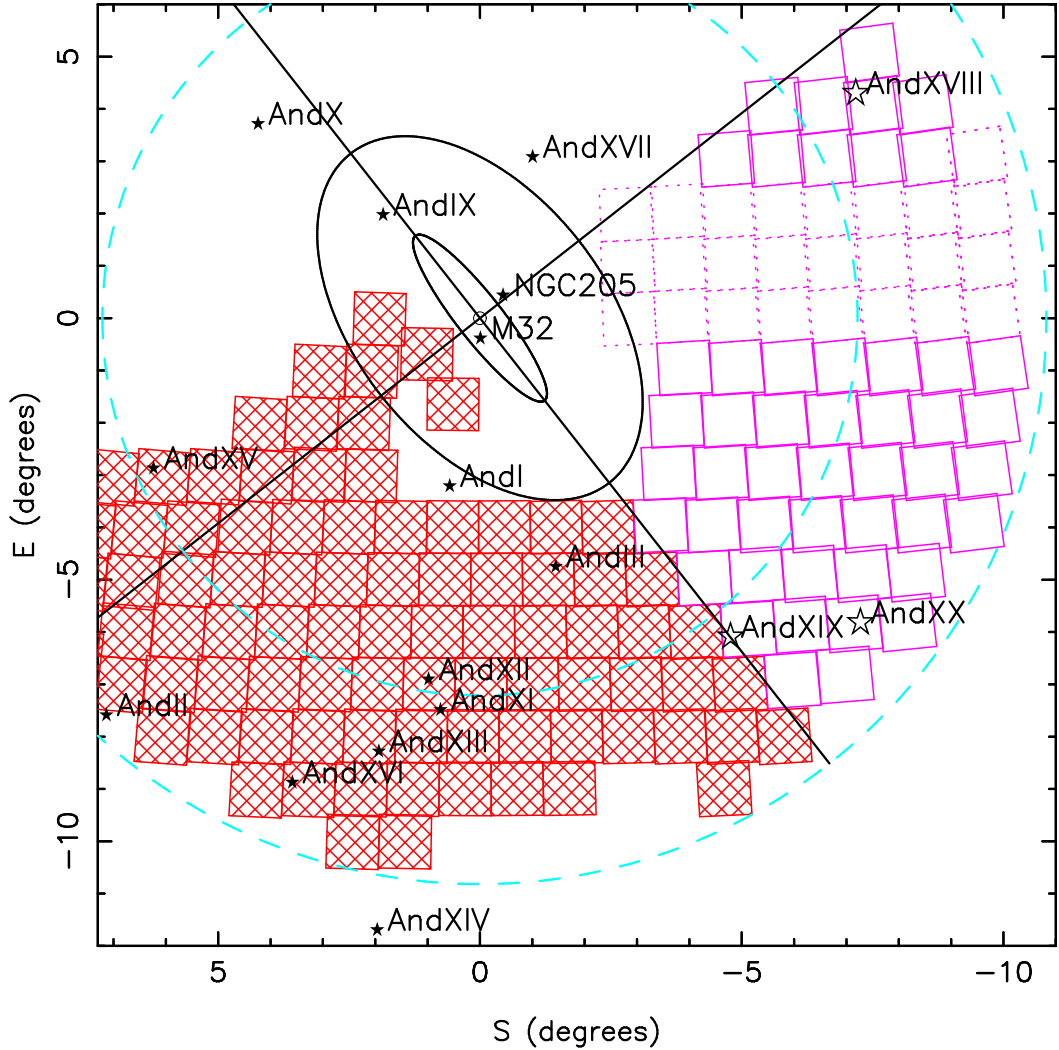}
    \caption{A tangent plane projection of the CFHT/MegaPrime survey
    area around M31. The inner ellipse represents a disk of
    inclination 77 degrees and radius 2 degrees (27\,kpc), the
    approximate edge of the regular M31 disk. The outer ellipse shows
    a 55\,kpc radius ellipse flattened to $c/a = 0.6$, the limit of
    the original INT/WFC survey (\citealt{ferguson2002}). Major and
    minor axes of M31 are indicated. The inner and outer blue dashed
    circles show maximum projected distances of 100\,kpc and 150\,kpc
    from the center of M31, respectively. Red hatched fields show the
    location of our extant imaging of the south-east quadrant of M31
    (\citealt{ibata2007}). Magenta fields show the location of fields
    for our ongoing survey of the south-west quadrant of M31 (solid
    lines denote observed fields, dotted lines denote fields still to
    be observed). Black stars show the locations of various known M31
    satellite galaxies, and open stars show the positions of the three
    new dwarf galaxies presented herein.}
  \end{center}
\end{figure}

\clearpage

\begin{figure*}
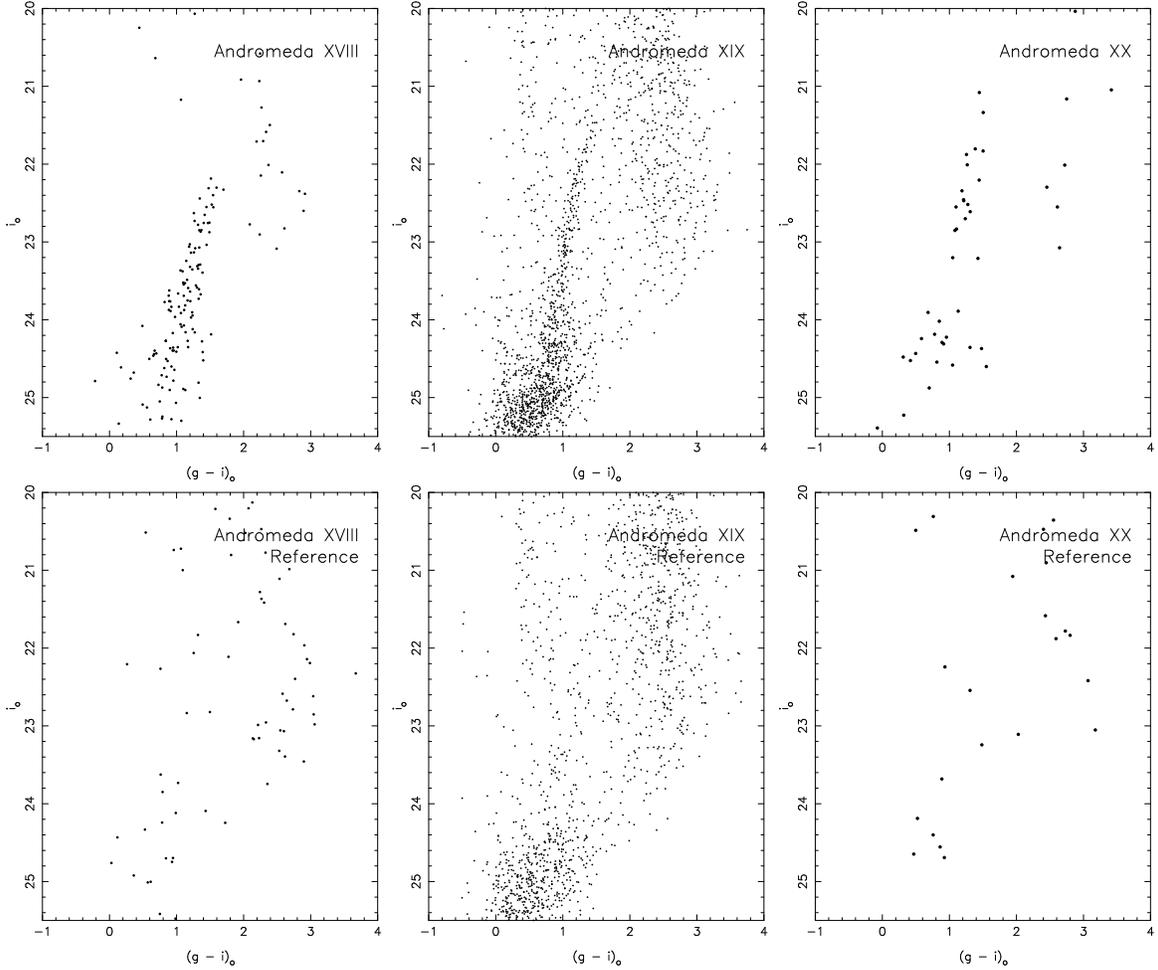

  \begin{center}
    \includegraphics[angle=270, width=5.cm]{f2a.ps}
    \includegraphics[angle=270, width=5.cm]{f2b.ps}
    \includegraphics[angle=270, width=5.cm]{f2c.ps}
    \includegraphics[angle=270, width=5.cm]{f2d.ps}
    \includegraphics[angle=270, width=5.cm]{f2e.ps}
    \includegraphics[angle=270, width=5.cm]{f2f.ps}
    \caption{The top panels show a comparison of the $i_o$ versus $(g
    - i)_o$ colour-magnitude diagrams of the three newly discovered
    Local Group galaxies, Andromeda~XVIII (left), XIX (middle) and XX
    (right), where all stars lying within 2 half-light radii from the
    center of each galaxy have been plotted (corresponding to 1.8,
    12.4 and 1 arcmins, respectively). The bottom panels are reference
    fields probing an equivalent area offset from each galaxy by
    several half light radii. A red giant branch (RGB) is visible in
    each galaxy, although in the case of Andromeda~XX it is very
    sparsely populated. None of the galaxies display any evidence for
    bright blue stars (either bright main sequence or blue-loop),
    indicative of a young population, and in this respect they
    resemble the typical stellar populations of dSph galaxies. The
    faint blue objects centered around $i_o \sim 25.2$ with a mean
    colour of $(g - i)_o \sim 0.5$ in the Andromeda~XIX CMD may be a
    horizontal branch component, although contamination from
    misidentified galaxies is considerable in this region of
    colour-magnitude space.}
  \end{center}
\end{figure*}
 
\clearpage

\begin{figure*}
  \begin{center}
    \includegraphics[angle=270, width=16.cm]{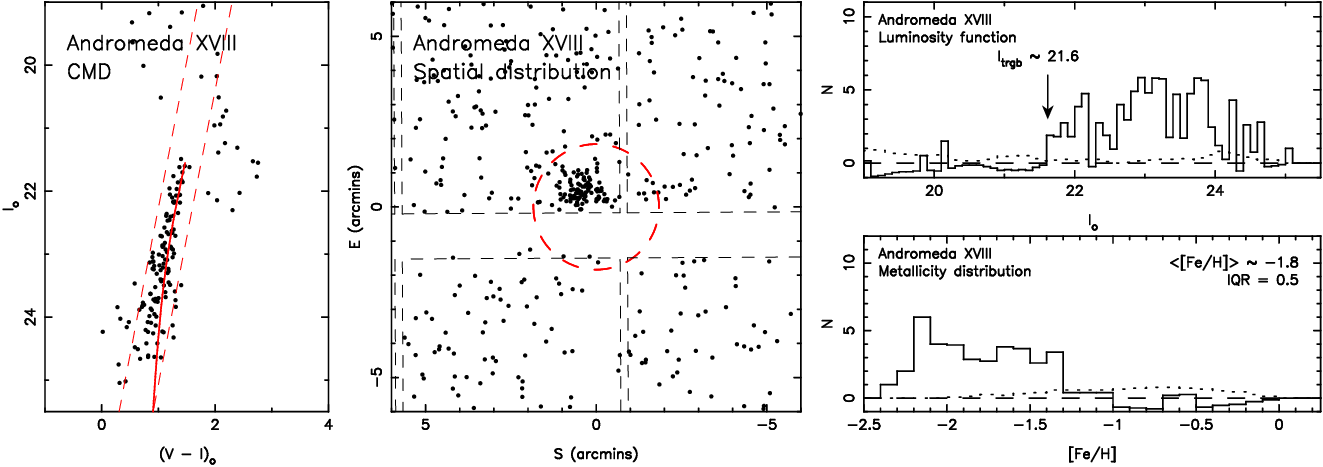}\vspace{0.5cm}
    \includegraphics[angle=270, width=16.cm]{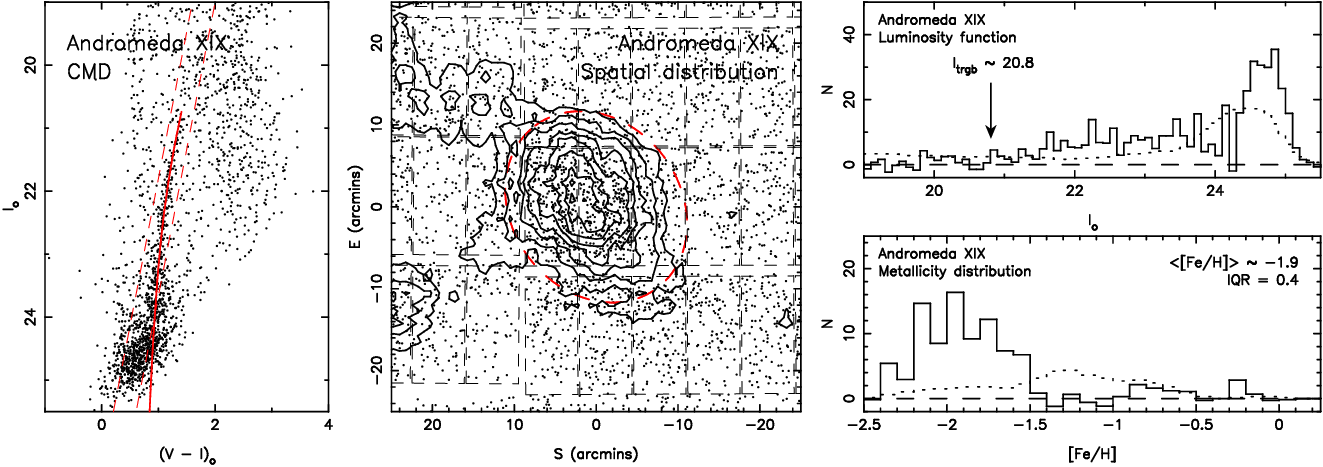}\vspace{0.5cm}
    \includegraphics[angle=270, width=16.cm]{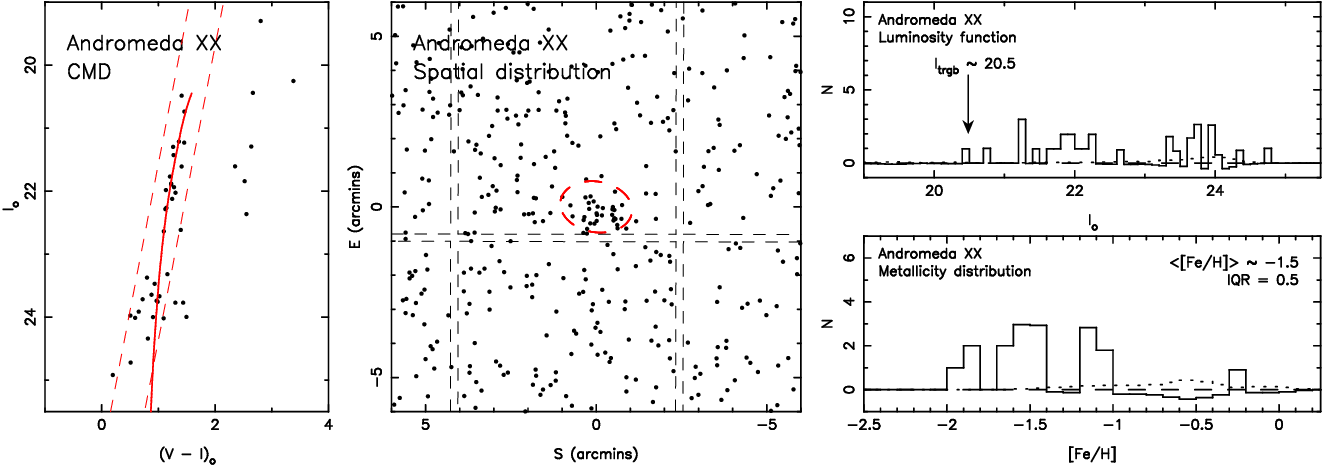} 
  \end{center}
\end{figure*}

\clearpage

\begin{figure*}
  \begin{center}
    \caption{Each set of panels shows various properties of
    Andromeda~XVIII (top row), Andromeda~XIX (middle row) and
    Andromeda~XX (bottom row). Left panels: $I_o$ versus $(V - I)_o$
    colour magnitude diagram for each galaxy. Dashed lines define a
    colour-cut used to preferentially select stars associated with the
    dwarf. A 13\,Gyr isochrone with the representative metallicity of
    the dwarf from \cite{vandenberg2006}, shifted to the appropriate
    distance modulus, is overlaid on each CMD. Only stars within the
    dotted ellipses shown in the middle panels are plotted. Middle
    panels: tangent plane projections of the spatial distribution of
    stars in the vicinity of each dwarf. Only stars satisfying the
    colour cuts shown in the CMDs are plotted. Dashed lines show the
    edges of the CFHT/MegaPrime CCDs.  Dashed ellipses mark two
    half-light radii from the center of each galaxy. For
    Andromeda~XVIII and XX, the dwarf galaxies are clearly visible as
    overdensities in the centers of each field, whereas Andromeda~XIX
    is more extended and diffuse and contours have been overlaid to
    more clearly define its structure. The first contour is set $3 -
    \sigma$ above the background, and subsequent contour levels
    increase by $1.5\,\sigma$ over the previous level. Right top
    panels: foreground-corrected, de-reddened, $I$-band luminosity
    functions of stars in each galaxy satisfying our colour and
    spatial cuts. Scaled reference field luminosity functions are
    shown as dotted lines. The estimated luminosity of the tip of the
    red giant branch is highlighted. Right bottom panels:
    foreground-corrected observed photometric metallicity distribution
    function derived using the technique detailed in
    \cite{mcconnachie2005a} using 13\,Gyr isochrones from
    \cite{vandenberg2006} with [$\alpha$/Fe] $= 0$]. Scaled reference
    field metallicity distribution functions are shown as dotted
    lines.  The mean metallicity and metallicity spread (quantified
    using the inter-quartile range, IQR) for each galaxy is
    highlighted.}
  \end{center}
\end{figure*}

\clearpage

\begin{figure}
  \begin{center}
    \includegraphics[angle=270, width=14.cm]{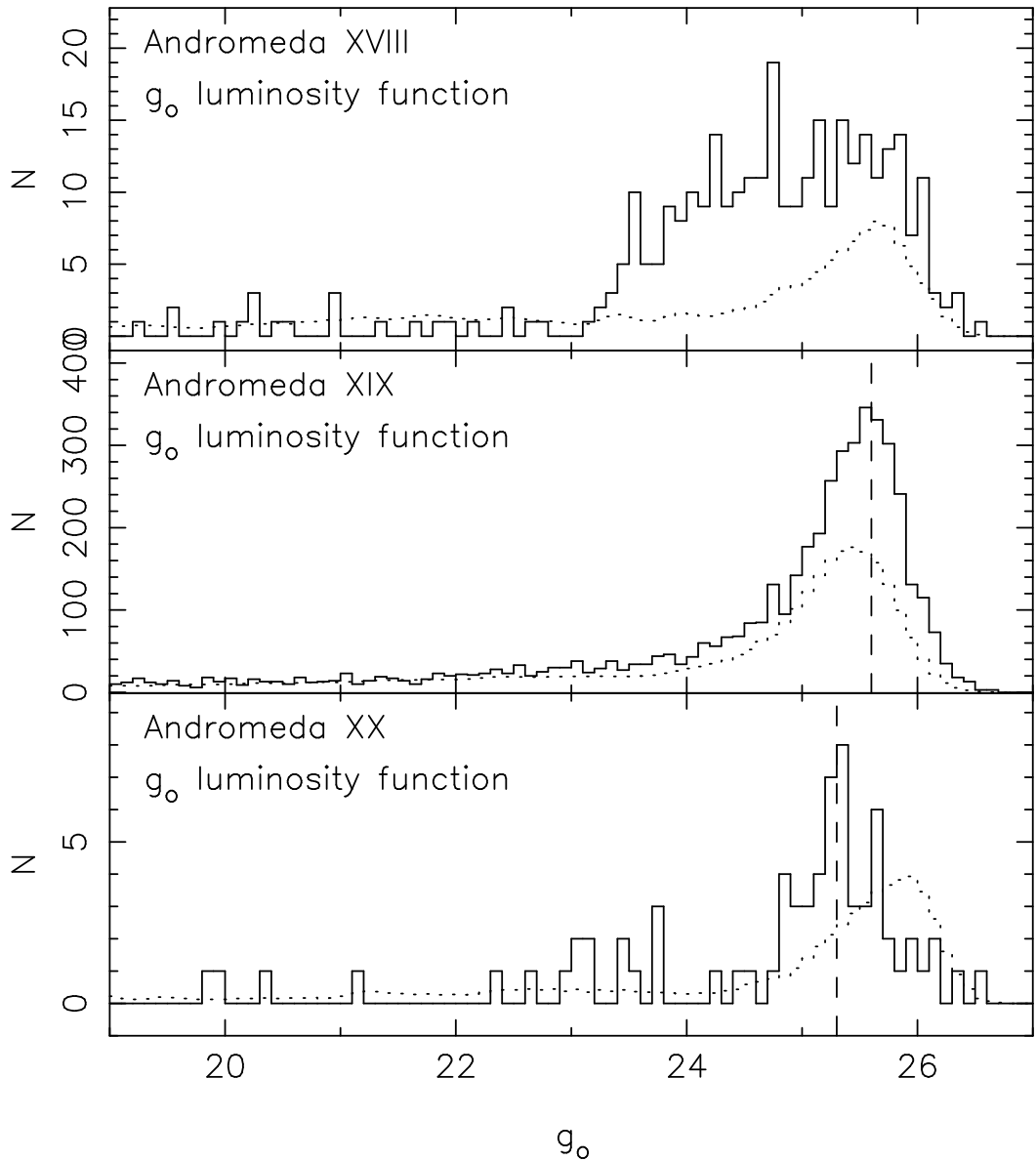}
    \caption{De-reddened $g$-band luminosity functions of stars within
    2 half-light radii of Andromeda~XVIII, XIX and XX (top, middle and
    bottom panels, respectively). These luminosity functions go deeper
    than the CMDs shown previously since only detection in the
    $g$-band is required. Nearby reference fields, scaled by area, are
    shown as dotted lines in each panel. For reference, the horizontal
    branch of M31 has a magnitude of $g_o \sim 25.2$. In Andromeda~XIX
    and XX, we attribute the peak of stars at $g_o \sim 25.3$ and $g_o
    \sim 25.6$, respectively, to a detection of horizontal branch
    stars (indicated by the dashed lines). In Andromeda~XVIII, no
    feature attributable to the horizontal branch is visible, as
    expected from its larger distance. Thus these detections (and
    non-detection) are consistent with the distances derived for these
    galaxies via the TRGB analysis.}
  \end{center}
\end{figure}

\clearpage

\begin{figure}
  \begin{center}
    \includegraphics[angle=180, width=9.cm]{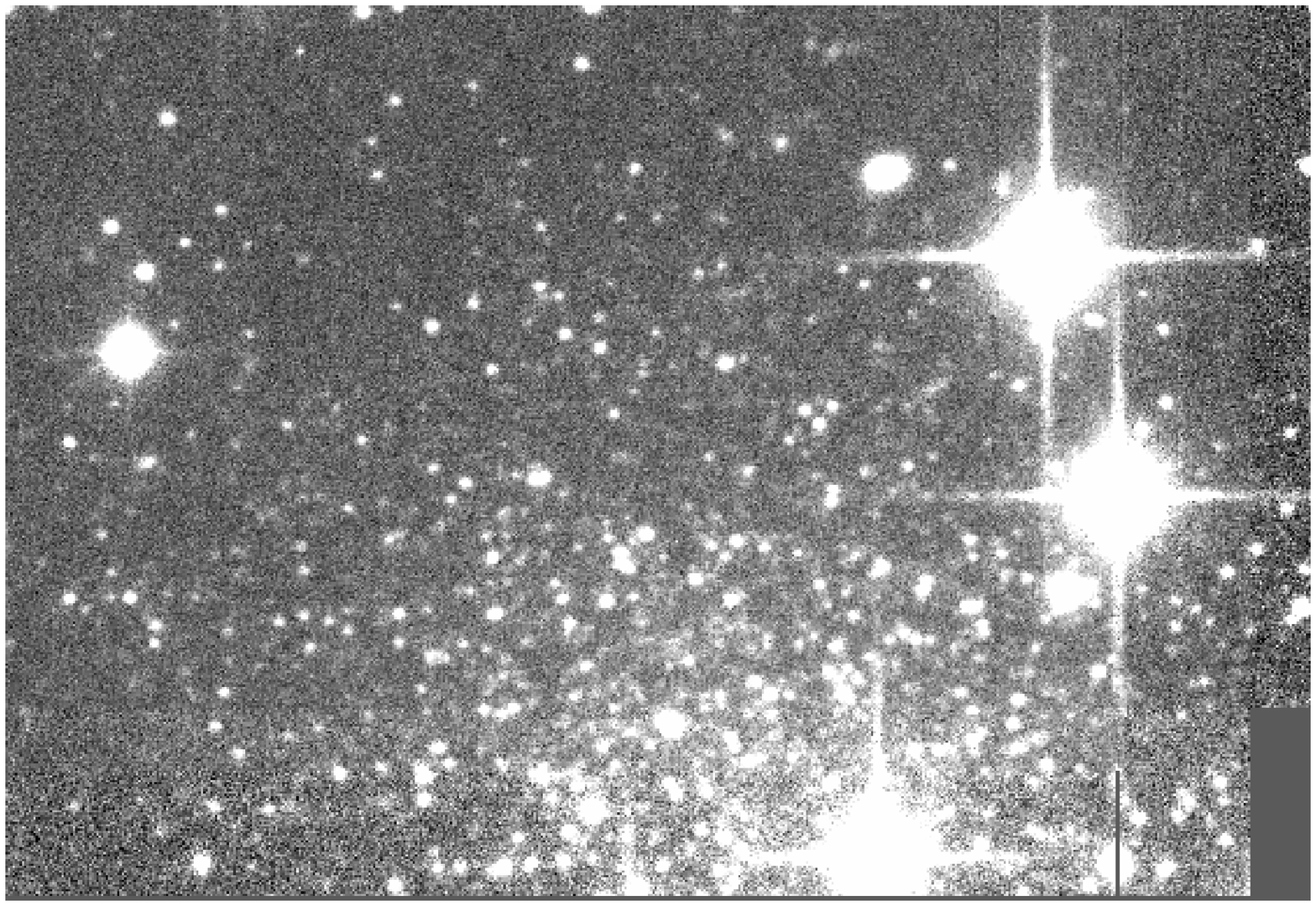}\vspace{1cm}
    \includegraphics[angle=180, width=9.cm]{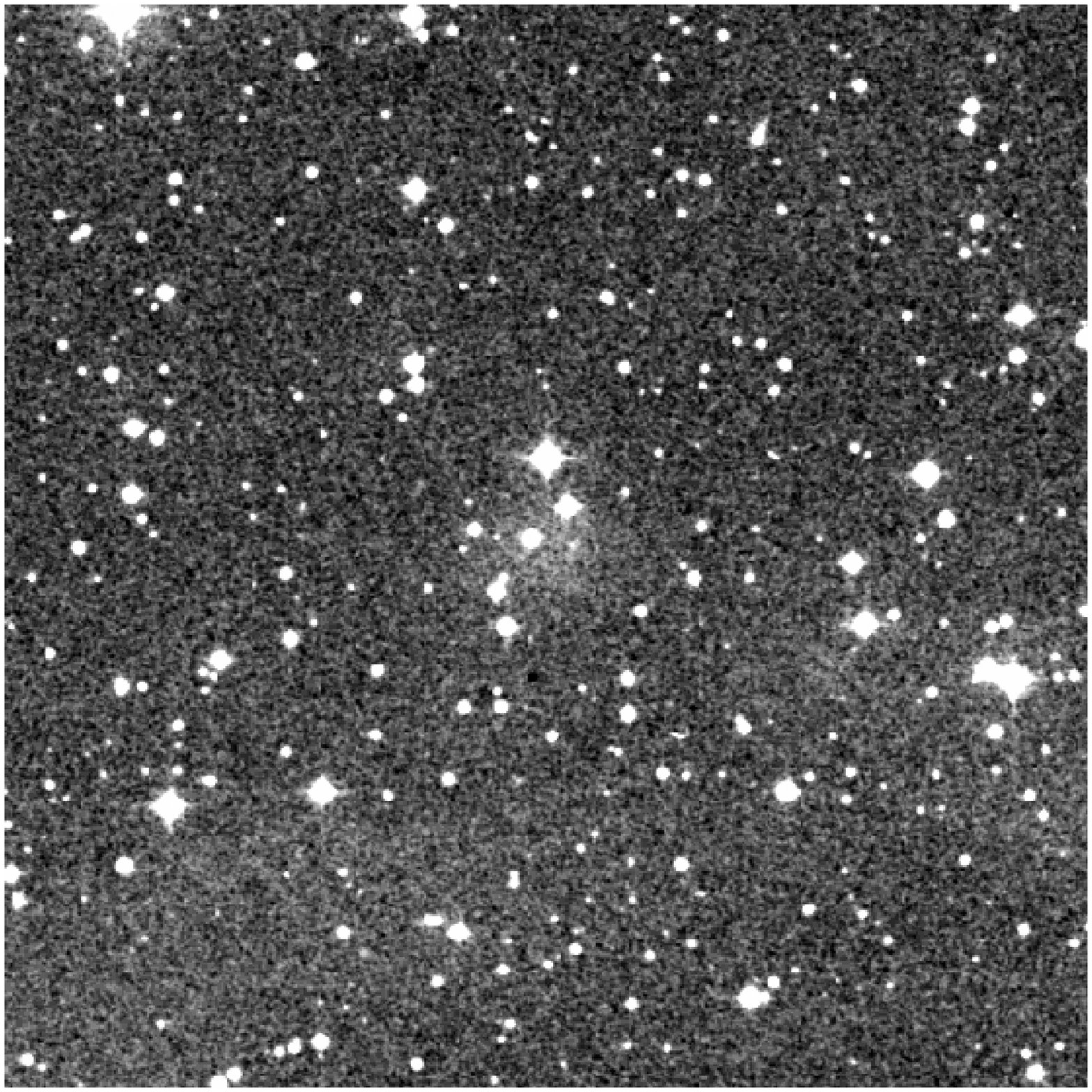}
    \caption{Top panel: The CFHT/MegaPrime $i-$band image of
    Andromeda~XVIII with linear scaling. Approximately $2.5 \times
    1.2$\,arcmins in the vicinity of Andromeda~XVIII is shown. This
    galaxy lies in the south-west corner of one of the CCDs, and some
    of it remains hidden behind the large gap between the second and
    first rows of CFHT/MegaPrime mosaic. Unlike the majority of recent
    discoveries in the Local Group, Andromeda~XVIII is clearly visible
    based upon its resolved light. Bottom panel: A $10 \times
    10$\,arcmins image centered on the coordinates of Andromeda~XVIII,
    with linear scaling, from the POSS2/UKSTU (Blue) survey, taken
    from the Digitized Sky Survey. Andromeda~XVIII is visible at the
    center. Some Galactic nebulosity is also present in this
    region. This galaxy is also visible on the original POSS1 (Blue)
    survey plates, and suggests that there may be other comparably
    bright galaxies within the Local Group which have so far eluded
    detection. In each panel, North is to the top and East is to the
    left.}
  \end{center}
\end{figure} 

\clearpage

\begin{figure*}
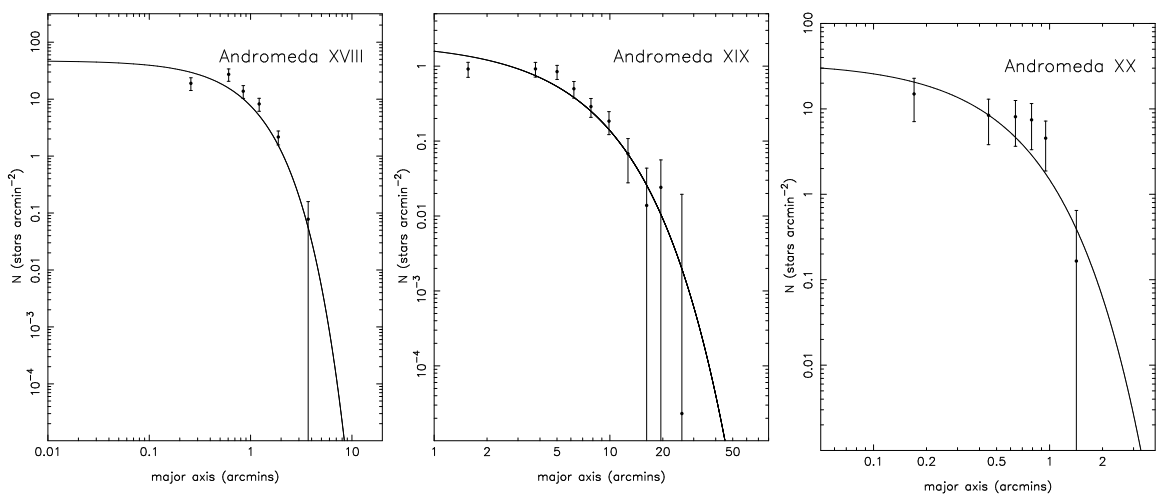

  \begin{center}
    \includegraphics[angle=270, width=5.cm]{f6a.ps}
    \includegraphics[angle=270, width=5.cm]{f6b.ps}
    \includegraphics[angle=270, width=5.cm]{f6c.ps}
    \caption{Radial profiles of Andromeda~XVIII, XIX and XX (left to
    right, respectively), derived in the same way as in
    \cite{mcconnachie2006b} using elliptical annuli with the position
    angles, ellipticities and centroids listed in Table~1. Overlaid on
    these profiles are the most probable exponential profiles derived
    using the same maximum likelihood technique.}
  \end{center}
\end{figure*} 

\clearpage

\begin{figure}
  \begin{center}
    \includegraphics[angle=270, width=14.cm]{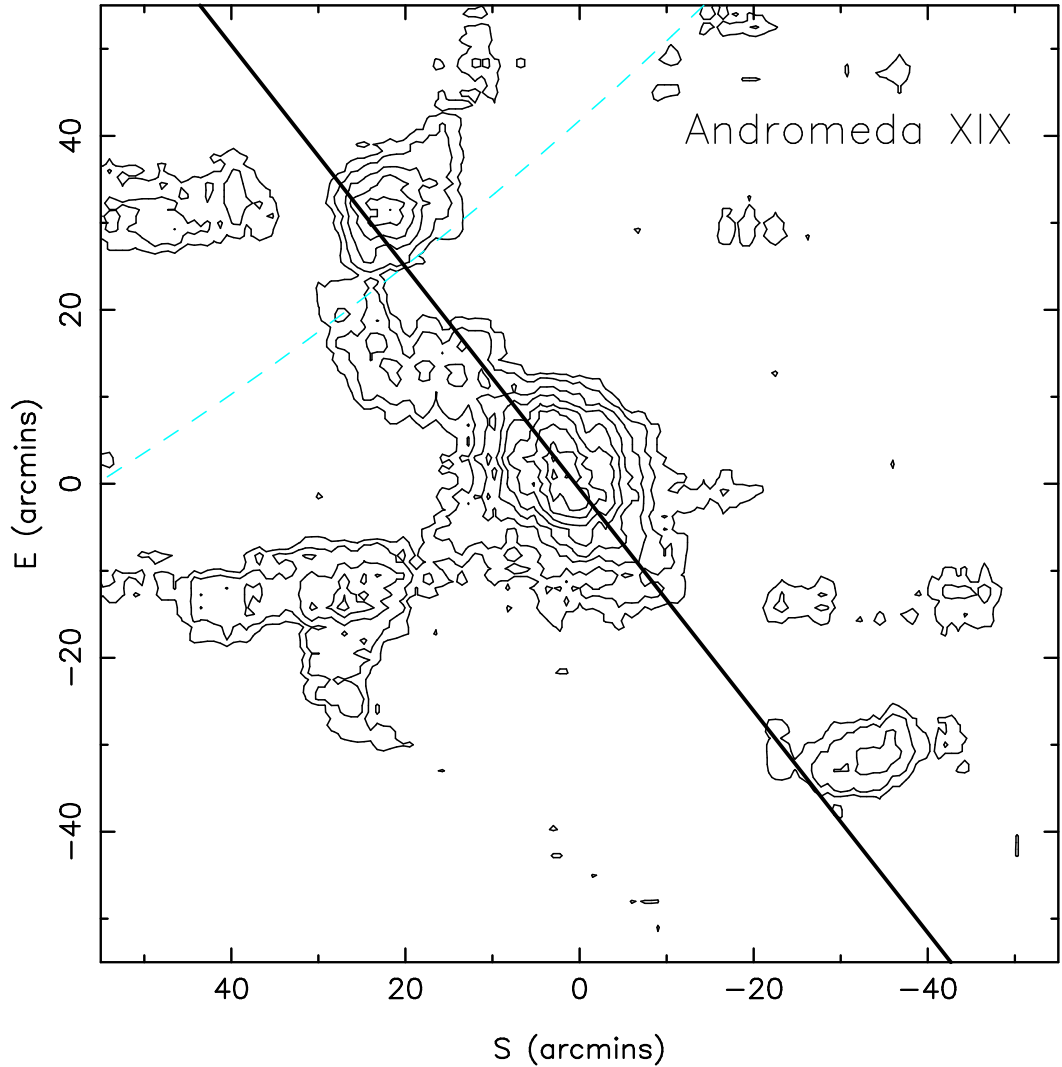}
    \caption{The surroundings of Andromeda~XIX, shown as a tangent
    plane projection of the stellar density. The first two contour
    levels are $2$ and $3 - \sigma$ above the background, and then
    they increase by $1.5\,\sigma$ over the previous level.  The major
    axis of M31 is shown as the solid black line, and the blue dashed
    line shows part of the circle which marks the $100$\,kpc boundary
    from the center of M31, as in Figure 1. Andromeda~XIX lies close
    to the major axis substructure identified in \cite{ibata2007},
    some of which can be seen in this plot. There is also some
    evidence of tidal features in the outskirts of Andromeda~XIX.}
  \end{center}
\end{figure}

\end{document}